\documentclass[12pt,a4paper,epsfig]{article}
\usepackage{epsfig}
\begin{document}

\author{{Dorodnitsyn A.V.}}
\title{\bf Line-driven winds in presence of strong gravitational fields}
\date{}
\maketitle

\bigskip
\noindent
Space Research Institute, \newline
117997,
Profsoyuznaya st. 84/32, \newline
Moscow, Russia\newline
(dora@mx.iki.rssi.ru)

\bigskip

\section*{Abstract}
\noindent We propose a general physical mechanism which could
contribute to the formation of fast line-driven outflows at
the vicinity of strong gravitational field sources. We argue that
the gradient of the gravitational potential plays the same role as
the velocity gradient plays in Sobolev approximation. Both Doppler
effect and gravitational redshifting are taken into account in
Sobolev approximation. The radiation force becomes a function of
the local velocity gradient and the gradient of the gravitational
potential. The derived equation of motion has a critical point
that is different from that of CAK. A solution that is continuous
through the singular point obtained numerically. A comparison with
CAK theory is presented. It is shown that the developed theory
predicts terminal velocities which are greater than those obtained
from the CAK theory.

\bigskip
\noindent
{\bf Key words:} hydrodynamics - quasars: general - radiation
mechanisms

\section*{Introduction}
Acceleration of matter due to radiation pressure in lines
plays an important role in formation of outflows from hot
stars and possibly from active galactic nuclei (QSO, Seyfert galaxes, etc). This
paper is the first in a series of papers on the effects that
strong gravitational fields play in formation and structure of
winds driven by radiation pressure in lines.

Theory of winds from O-type stars is well-developed and has
a good agreement with observations.
In a pioneering paper of Sobolev (1960),
it was recognized that a
radiation transfer in accelerated medium is simplified drastically
in comparison with that of static atmosphere.
The importance of the line
opacity for the formation of winds from hot stars was pointed out
in paper of Lucy \& Solomon (1970).
A prominent step
in this field was made in papers of Castor, Abbott \& Klein (1975),
( hereafter
CAK ). In these papers it was shown that absorption of the
radiation flux in lines can be an effective mechanism
of redistributing of momentum from radiation to the wind.
The presence of the velocity gradient gives additional
effect on the acceleration because of the Sobolev effect.
CAK showed that the resultant radiation force, which is due to
absorption in an ensemble of optically thin and optically thick
lines, may be several orders of magnitude greater
than that due to electron scattering.
The ideas and technique of CAK were developed by
Abbott (1980) and in works of many other authors.

Theory of winds accelerated by the radiation pressure in lines
is usually applied to explain outflows from active galactic nuclei (AGN).
The many puzzling, observational  characteristic features of AGNs
are: the blue-shifted, relative to the emission line rest frequency
broad absorption lines (BALs) - the most convincing evidence
of outflows with velocities as large as $0.1 c$; NALs -narrow
absorption lines seen in UV and X-ray spectra associated with
$\sim 1000  \,{\rm km\cdot s^{-1}}$ outflows;
BELs - broad emission lines observed in UV indicate the flow with
the characteristic velocity about several thousand ${\rm km\cdot s^{-1}}$.
From the short  time-scale of the X-Ray variability (Tennant {\it et al} 1981)
it is concluded that
the size of the emitting region is $\sim 10^{14}{\rm cm}$.
(Although there exist an uncertainty in the estimation of the
column densities, see Arav et al, 2002, and references therein)
Only black hole (BH)
as a central
object can couple this small length scale with the total
luminosity  $L\sim 10^{46} {\rm egr\cdot s^{-1}}$. The existence
of BALs together with large luminosity makes radiation driven
mechanism plausible. Line-locking, observed in spectra of some QSO
 gives additional evidence of the importance of radiation in
acceleration of the wind.
Murray at al. (1995) model the wind that originates not far from BH $\sim 10^{16} {\rm cm}$.
Simplifying assumptions allowed to solve separately equation of
motion in radial and polar angle directions. Two-dimensional
time-dependent hydrodynamic calculations of radiation-driven winds
from luminous accretion disks were made in
Proga et al. (1998), Proga et al. (2000).

In all papers where the radiation pressure was assumed as a main
driving force forming outflows from AGNs, authors adopted modifications of the CAK
theory (although effects which are not included in the CAK theory were taken into account).
A serious
difference between a hot star wind and AGN outflow is due to different
geometry. Wind from O-type star is nearly spherically-symmetric
whereas an outflow from AGN originates from accretion disc an thus
is closely axially-symmetric. From observations we know that wind
from AGN is exposed to a very high X-ray and UV flux.
The AGN spectra is harder then that of O-type star thus the
radiation force could be smaller due to a higher ionization
state. On the other hand the observations of spectral lines from
moderately and highly
ionized species pose a problem of how the wind avoids
over-ionization. This problem can be overcame if there exist some
gas that shields the wind from the ionizing radiation. The second
way is to assume that the wind is in the form of dense clouds
which are confined by some mechanism. The ionizing effects of the
X-ray flux in the winds from massive X-ray binary systems
were considered in Stevens \& Kallman (1990).

The main underpinning
assumption of CAK, is that of the velocity gradient leads to enhancement
of the radiation force. If there is no velocity gradient
then the radiation force is simply due to absorption of
the fraction of the radiation flux which is blocked by lines.
If there is a velocity gradient, the absorbing ions are exposed
to the unattenuated radiation flux due to Doppler shifting.

In this paper we propose a general physical mechanism which
could play an important role in the formation of fast
outflows at the vicinity of strong gravitational field sources.
We argue that the gradient of the gravitational potential
$d\phi/dr$ (in case of sufficiently strong
gravitational fields) plays the same role as $dv/dr$
plays in acceleration of line-driven winds from hot stars.

Taking into account
gravitational redshift of frequency
we conclude that the gradient of the gravitational potential
allows a line to shift out from behind of its own shadow. The
line is shifted to the extent where the continuum flux is not reduced by
absorption, hence $d\phi/dr$ exposes the line to powerful radiation.
Thus we call the resultant wind: "Gravitationally Exposed Flow" (GEF).

In this paper the following toy model is adopted. Consider a wind
accelerated by the radiation pressure in lines at the vicinity of
BH. For simplicity we assume the wind to be spherically-symmetric
and skip ionization problem. These crude assumptions are
justified by the question we are going to answer
that is to compare a solution for gravitationally exposed flow
with CAK solution. We are addressing 2d
modeling of the problem to the separate paper.

The plan of the paper is as follows. In Section 1. we discuss
relations for the optical depth $\tau_L$ and radiation force
$f_r$ in standard CAK theory. In Section 2. a relation for the radiation force is derived
taking into account gravitational redshifting. Equation of motion
describing stationary, spherically-symmetric wind is derived.
Properties of the solution at the critical point are investigated. The numerical
solution is provided. The results are summarized in
the Discussion and Conclusions.

\section{Radiation force}
Assume that a radiation flux $F_c (\nu)$
($F_c(\nu_i)$ - is the continuum flux per interval of frequency
on the line frequency $\nu_i$)
emitted by the photosphere (or disc) is
radially-directed. The radiation force that results from
absorption in a single line with the Doppler width
$\Delta\nu_D\simeq \nu v_{th}/c$ may be
represented in the following form (Castor, Abbott \& Klein 1975):

\begin{equation}\label{force0}
f_l=\kappa_l \left(\frac{\Delta\nu_D F_c(\nu_i)}{c}\right)
\frac{1-e^{-\tau_l}}{\tau_l}
\mbox{,}
\end{equation}
The effective optical depth $\tau_l$ is calculated on the frequency of
line, which obtained taking into account the Doppler shifting,
$\kappa_l$ ($\rm cm^2\,g^{-1}$) is the opacity at the line centre:
\begin{equation}\label{kappa}
  \kappa_l=\frac{\pi e^2}{m c} \frac{g f}{\rho}
  \frac{N_l/g_l-N_u/g_u}{\Delta\nu_D}\mbox{.}
\end{equation}
In Sobolev approach $dv/dr$ is taken into account when
calculating $\tau_l$.
A relation for the optical depth was derived by Castor (1970):
\begin{equation}\label{taucak}
  \tau_l=\int_r^\infty \psi\rho\kappa_l \,dr
  \simeq \frac{\rho v_{th} \kappa_L}{|dv/dr|}\mbox{.}
\end{equation}
When calculating $\tau_l$ it is necessary to take
into account only those absorbers which are in a
section of the column across which the velocity changes by
$v_{th}$.
Relation ~(\ref{taucak}) was obtained assuming that the line
profile $\psi$ is a delta function.
CAK introduced an optical depth parameter independent of line
opacity:

\begin{equation}\label{taut}
t\equiv\sigma_e\rho\frac{v_{th}}{dv/dr}\mbox{,}
\end{equation}
where $\sigma_e=0.4\, {\rm cm^2\cdot g^{-1}}$. If the velocity
gradient is large the optical depth parameter $t$ may be much
less then the corresponding electron optical depth.
If $t\sim 1$ the radiation force is determined
by the part of the radiation flux that is blocked by lines. In
the opposite case of small $t$, the radiation force may
exceed gravity by several orders of magnitude for a typical O-star atmosphere.
Summing equation ~(\ref{force0}) over an ensemble of lines, CAK
obtained the following relation:
\begin{equation}\label{rforce}
f_l=\frac{L\sigma_e}{4\pi r^2 c} M(t)\mbox{,}
\end{equation}
where $M(t)$ is the force multiplier:
\begin{equation}\label{Mt}
  M(t)=\sum_i{\frac{F_c(\nu_i)\Delta\nu_D}{F}}\min(\beta_L,t^{-1})\mbox{,}
\end{equation}
where $\beta_L=\kappa_L/\sigma_e$ (note that $t=\tau_L/\beta_L$),
$F$ - is the total radiation flux.
$\tau_l/\kappa_l$ - is the column of
mass which absorbs radiation, $(1-e^{\tau_l})$
multiplier in ~(\ref{force0}) gives the probability of absorption.
If there are only strong lines then the radiation force is
proportional to the number of strong lines. Thus in optically thick case
$f_l$ is independent of the line strength and $f_l\sim dv/dr$.
If lines are
all optically thin then each ion is absorbing the unattenuated
radiation flux, and thus the radiative acceleration is independent
of the wind dynamics. For the ensemble of optically thin and
thick lines CAK found that $M(t)$ can be fitted by the power-law:
\begin{equation}\label{multipl}
  M(t)=kt^{-\alpha}\mbox{.}
\end{equation}
$\alpha=1$ represents a case when all lines are optically thick,
$\alpha=0$ - optically thin. With ~(\ref{multipl}) and
~(\ref{taut}), equation ~(\ref{rforce}) reads:

\begin{equation}\label{force01}
f_L=\frac{L\sigma_e k}{4\pi r^2 c}
(\sigma_e v_{th})^\alpha \left(\frac{1}{\rho}\frac{dv}{dr}\right)^\alpha \mbox{.}
\end{equation}

\section{Gravitationally Exposed Flow}
\subsection{{\bf The radiation force due to gradient of gravitational
field}}

\smallskip
\noindent
If a wind is
accelerated in a strong gravitational field we may
expect that a photon emitted by the disc may become resonant with
the opacity of some line not only because of the Doppler effect
but also because of the gravitational redshifting.
The redshifting of the photon's frequency is the only effect of general relativity
(hereafter GR) we take into account in our approximate treatment.
A photon with the frequency $\nu_d$, emitted at the point with
the gravitational potential $\phi_1$ will be registered at the
point where $\phi=\phi_2$ ($\mid\phi_1\mid>\mid\phi_2\mid$, $\phi<0$)
reddened
according to the well-known approximate formulae (see for example Landau \& Lifshitz, 1960):

\begin{equation}\label{redden1}
\Delta\nu\equiv\nu_2-\nu_d=\frac{\phi_1-\phi_2}{c^2}\nu_d\mbox{.}
\end{equation}
If there is a velocity difference between these two points the photon will
be additionally red-shifted due to the Doppler effect. The
resultant frequency, seen by the absorber at his rest frame reads:
\begin{equation}\label{frequency1}
  \nu\simeq\nu_d+\nu_d\left(-\frac{v}{c}+\frac{\Delta\phi}{c^2}\right)\mbox{.}
\end{equation}
To derive a relation for the radiation force we need to calculate the corresponding
Sobolev optical depth, but now taking into account the
gravitational redshifting. Assuming that absorption occurs in a
line with the delta-function profile, for the optical depth in a
radial direction we obtain:

\begin{equation}\label{optdepth}
  \tau_l=\int_r^\infty \kappa_l \rho \,dr\simeq
  \frac{\rho v_{th} \kappa_l}{dv/dr
  +\frac{1}{c}d\phi/dr}\mbox{.}
\end{equation}
A characteristic length: $\delta r\sim
v_{th}/(dv/dr+\frac{1}{c}d\phi/dr)$ gives a thickness of a shell where
the absorption due to a single line take place.

In our
calculations we are neglecting the influence of the gravitational
field on the geometry of space-time, because in that case it our
treatment will be too complicated.
Introducing an optical depth parameter, which is
analogous to ~(\ref{taut}), we have:

\begin{equation}\label{taut2}
t\equiv\frac{\sigma_e\rho v_{th}}{dv/dr+\frac{1}{c}d\phi/dr}\mbox{.}
\end{equation}

Note that relations ~(\ref{rforce}), ~(\ref{multipl}) are derived
in such a way that all information about wind dynamics
(distribution of $v$ that is responsible of redshifting)
is imbedded into parameter $t$.
It is assumed that $\tau$ must be
calculated taking into account redshifting. No assumptions about
the particular physical mechanism that produces redshifting are
made. In our case $t$ contains
also an information about additional redshifting which is due to
gradient of gravitational potential. Note that ~(\ref{taut2}) may be represented in the form
${\displaystyle t=\frac{\sigma_e\rho
v_{th}}{dv^\star/dr}}$, where ${\displaystyle v^\star=
v+\frac{1}{c}\phi}$. The distribution of $v^\star$
contains all the information about redshifting.
These considerations hold so far Sobolev approximation is
assumed and until the influence of the gravitational field on the
geometry of space-time is neglected. We are justified
in making use of ~(\ref{taut2}) instead of ~(\ref{taut}) when
calculating ~(\ref{multipl}).

By substituting
~(\ref{taut2}) into ~(\ref{rforce}) and ~(\ref{multipl})
the radiation force is obtained:

\begin{equation}\label{force1}
f_L=\frac{L\sigma_e k}{4\pi r^2 c}
\left(\frac{4\pi}{|{\dot M}|\sigma
v_{th}}\right)^\alpha\left[vr^2\left(\frac{dv}{dr}+\frac{1}{c}
\frac{d\phi}{dr}\right)\right]^\alpha\mbox{,}
\end{equation}
where the radiation force was transformed making
use of the continuity equation:

\begin{equation}
\mid {\dot M}\mid=4\pi\rho v r^2\mbox{.}
\end{equation}

\noindent
\subsection{{\bf Basic equations}}

\noindent
\smallskip
An equation of motion describing stationary,
spherically-symmetric, isothermal, line-driven wind reads:

\begin{equation}\label{momentum1}
v\frac{dv}{dr}=-\frac{1}{\rho}\frac{dP}{dr}-\frac{d\phi}{dr}+
\frac{L\sigma}{4\pi r^2 c}+
\frac{L\sigma_e k}{4\pi r^2 c}
\left(\frac{4\pi}{|{\dot M}|\sigma
v_{th}}\right)^\alpha\left[vr^2\left(\frac{dv}{dr}+\frac{1}{c}
\frac{d\phi}{dr}\right)\right]^\alpha\mbox{.}
\end{equation}
As well as the equation of motion of CAK, equation ~(\ref{momentum1}) is nonlinear
with respect to the velocity gradient. This bahaviour complicates a structure
of ~(\ref{momentum1}) and slows numerical solution. Investigating
~(\ref{momentum1}) we hope to find a trans-critical solution,
which starts subsonically, proceeds through the critical point and
goes to infinity approaching a terminal velocity $v^{\infty}$.
Equation ~(\ref{momentum1}) resembles CAK equation of motion
except the radiation pressure term. Instead of $(dv/dr)^\alpha$ term,
equation ~(\ref{momentum1}) includes a combination:
$(dv/dr+\frac{1}{c}d\phi/dr)^\alpha$. As we will see this property of the equation
~(\ref{momentum1})
changes the structure of transcritical solution.

We adopt two types of potentials: Newtonian
potential (NP) and Paczynski and Wiita (hereafter PW ) potential:

\begin{equation}\label{poten1}
\phi=-\frac{GM}{r-r_g}\mbox{.}
\end{equation}
We make use of the modified potentials to model approximately the
GR effects.
PW potential, for the Schwarzschild black hole correctly
reproduces the positions of both the last stable orbit and the
marginally bound orbit. The many properties of modified potentials
are discussed in Artemova, Bjornsson \& Novikov (1996).
Introduction of the modified
potentials
allows to take approximately into account some properties of the
exact relativistic description. For simplicity hereafter we use Newtonian
potential. The case of the PW-potential is described in Appendix.
For Newtonian gravity $\phi=-GM/r$, equation ~(\ref{momentum1})
reads:

\begin{eqnarray}
F&=&\left(1-\frac{a}{v^2}\right)v\frac{dv}{dr}-\frac{2a^2}{r}+
\frac{GM}{r^2}-\frac{L\sigma}{4\pi r^2 c}\nonumber\\
&-&\frac{L\sigma k}{4\pi r^2 c}
\left(\frac{4\pi}{|{\dot M}|\sigma
v_{th}}\right)^\alpha\left[vr^2\left(\frac{dv}{dr}+\frac{1}{c}
\frac{GM}{r^2}\right)\right]^\alpha=0\mbox{,}\label{eqn1}
\end{eqnarray}
where ${\displaystyle a=\left(\frac{\partial P}{\partial \rho}\right)^{1/2}_T}$
- a
sound velocity. When obtaining equation ~(\ref{eqn1}) it was
assumed that the wind is isothermal $T={\rm const.}$ and
$P=a^2\rho$.

Introducing nondimmensional variables:
\begin{equation}
  x=\frac{r}{r_c}\mbox{,}\qquad \tilde v =\frac{v}{v_c}\label{nondimvar}
  \mbox{,}
\end{equation}
where $r_c$ is the critical point radius and $v_c$ is the velocity of matter
at the critical point. Taking into account ~(\ref{nondimvar}), equation ~(\ref{eqn1}) reads:

\begin{eqnarray}
F(p,v,x)&=&\left(1-\frac{a^2_1}{v^2}\right)v x^2 p-2a_1^2 x+
\frac{\beta^2}{2}\left(1-\Gamma\right)\nonumber\\
&&-\mu \beta^2\left[\frac{v}{\beta^2}x^2\left(p+
\frac{\beta}{2x^2}\right)\right]^\alpha=0\mbox{,}\label{dmless1}
\end{eqnarray}
where ${\displaystyle p\equiv \frac{dv}{dx}}$ and
${\displaystyle \Gamma\equiv\frac{L\sigma_e}{4\pi GM c}}$.
For simplicity hereafter we omit tilde.
The following nondimmentional parameters were introduced:

\begin{equation}\label{param}
a_1=\frac{a}{v_{c}}\mbox{,}\quad\beta=\frac{c}{v_{c}}\mbox{.}
\end{equation}
Note that ${\displaystyle \beta=\frac{c}{a} a_1}$ and there is only one independent parameter $a_1$.
A constant $\mu$ is determined according to relation:

\begin{equation}\label{mu1}
  \mu\equiv\Gamma k \left[\frac{8\pi GM}{\mid\dot M\mid\sigma_e
  v_{th}}\right]^\alpha
\end{equation}

Equation ~(\ref{dmless1}) is non linear with respect to $p$. The point where the speed of sound
is equal to velocity of the flow is no longer the singular point of the equation of motion.
To treat such equations a special technique must be applied.
\noindent
We are interested of the solution that starts subsonic near BH
is continuous through the singular point
and goes supersonically to infinity.
Singular point is defined by the condition:

\begin{equation}\label{singcond}
\left(\frac{\partial F}{\partial p}\right)_{c}=0\mbox{.}
\end{equation}
The second condition comes from the fact that at the critical point the velocity gradient is
continuous which requires that ${\tilde v}''$ is defined at the critical point.
The regularity condition reads:
\begin{equation}\label{regcond}
  \left(\frac{\partial F}{\partial x}\right)_{c}+
  p_c\left(\frac{\partial F}{\partial v}\right)_{c} =0\mbox{,}
\end{equation}
where ${\displaystyle
p_c=\frac{r_g}{v_{c}}\left(\frac{dv}{dr}\right)_{c}}$.
\newline
From ~(\ref{dmless1}),~(\ref{singcond}) the following useful relation is obtained:

\begin{equation}\label{mu}
\mu=\alpha^{-1} (1-a_1^2)\left[\frac{x_c^2}{\beta^2}\left(p_c
+\frac{\beta}{2x_c^2}\right)\right]^{1-\alpha}\mbox{,}
\end{equation}
where ${\displaystyle x_c=\frac{r_{c}}{r_g}}$. Substituting
~(\ref{dmless1}) to ~(\ref{singcond}), taking into account
~(\ref{mu}) will result in the following equation:

\begin{equation}\label{meq1}
2a_1^2 x_c^2 p_c^2+\frac{1}{2}(a_1^2-1)\beta p_c-2a_1^2=0\mbox{.}
\end{equation}
The momentum equation ~(\ref{dmless1}) is valid a hole
domain of the investigation, thus calculating it at the critical point and substituting
$\mu$ from ~(\ref{mu}) will obtain the following equation:

\begin{equation}\label{meq2}
\frac{\alpha-1}{\alpha} (1-a_1^2)x_c^2 p_c-2a_1^2 x_c+\frac{\beta^2(1-\Gamma)}{2}-
(1-a_1^2)\frac{\beta}{2\alpha}=0\mbox{.}
\end{equation}
Equations ~(\ref{meq1}),~(\ref{meq2}) form the system of equations
from which $p_c$ and $a_1$ may be calculated.
For example, assuming that the critical point is situated at $100\,r_g$, from
~(\ref{meq1}),~(\ref{meq2}) we find: $a_1=0.0393124$,
$p_c=10.4831$. As it will be clear
from the following, equations ~(\ref{meq1})-~(\ref{meq2}) are not
only more complicated than those of CAK, but also manifest the
different physical behavior of GEF. Equation ~(\ref{meq1}) differs
from those of CAK by the second term. Dropping in ~(\ref{meq1})
this term will reveal the result of CAK for the isothermal wind:
${\displaystyle p_c=\pm1/x_c^2}$, or ${\displaystyle
\left(\frac{dv}{dr}\right)_{c}= \pm\frac{v_{c}}{r_{c}} }$.
To solve ~(\ref{meq1}),~(\ref{meq2}) for $a_1$, $p_c$, we must take
into account that ${\displaystyle \beta=\frac{c}{v_{th}}a_1}$,
where ${\displaystyle v_{th}=\sqrt{k\frac{T}{m_n}}}$.

\subsection{Numerical solution}
For a given position of the critical point $x_c=r_{c}/{r_g}$
equations ~(\ref{meq1}),~(\ref{meq2}) are used to obtain
the velocity gradient at the critical point.
In contrast to CAK wind, the GEF case is more difficult to
analyze. In the isothermal limit of the CAK wind, equations
~(\ref{meq1}), ~(\ref{meq2}) could be solved analytically.
Anfortunately it is not possible (albeit see the end of this section)
in GEF case. In order to obtain $p_c$ and $a_1$ equations ~(\ref{meq1}),
~(\ref{meq2}) must be solved numerically.
All calculations in this paper are made for
$10^7 M_{\odot}$ black hole, the temperature of the flow is
$T=4\cdot 10^4 {\rm K}$ other parameters are: $\Gamma=0.5$, $\alpha=0.6$, $k=0.1$.

 In case of stellar winds a photospheric boundary
condition is usually adopted. Adjusting the position of the
critical point one obtains a solution that gives the position of
the photosphere ($r=r_{ph}$ at $\tau_e=2/3$) at some prescribed
radius $R$, which is identified with the radius of a star
(Bisnovatyi-Kogan 2001). Similar procedure was adopted by CAK.
In case of AGN such "photospheric"
conditions are clearly unphysical. The solution for the wind should be
continuously fitted with the solution for the accretion disc. This requires
2D modelling which is beyond the scope of this paper. In spherically-symmetric
approximation adopted here it is not
possible to fit self-consistently a wind solution with that of
accretion disc.

To compare GEF solution
with that of CAK we start deeply subsonic ($v\ll v_s$), starting from some initial density
$\rho_{in}$ and calculating both CAK and GEF solutions. Mathematically it is
equivalent to the problem of fitting of the wind solution with that of a static core
when calculating a structure of a star when mass loss is taken into account.
In such a case a
solution for the outflowing envelope is continuously fitted with
that of a static core.
As it results from a stellar wind theory
(for stationary spherically-symmetric wind) to fit continuously a
solution for a stationary outflowing envelope with that of a static core,
only $\rho$ and $T$ should be fitted. In case of stellar wind this
condition reads: $\rho_{in}^{env}=\rho_{out}^{core}$,
$T_{in}^{env}=T_{out}^{core}$ where the
fitting point must lay in a deep subsonic region $v\ll v_s$ (see
Bisnovatyi-Kogan \& Dorodnitsyn 1999 and references therein).

\noindent
The adopted procedure is equivalent to that if we start from
some static configuration ($v<<v_s$) with the density $\rho_{in}$ and then relax to stationary
wind solution (CAK and GEF).
Prescribing $\rho_{in}$ at the very bottom of the wind ($v\ll
v_s$) at $r_{in}$ we may find the solution which satisfies boundary condition.
This requirement allows to compare GEF solution with that of
obtained from CAK theory.

\noindent
We step out from the critical point $x_c$ by means
of the approximate formulas: $v=1\pm\mid p_c \mid \Delta x$, where
$\Delta x=x-x_c$. On every integration step equation
~(\ref{dmless1}) is solved numerically in order to obtain $dv/dx$.
The forth order, adaptive Runge-Kutta scheme was used to obtain
solutions depicted on Figure 1. - Figure 4.

Adjusting $x^{\rm CAK}_{c}$, $x^{\rm GEF}_{c}$ we
select those solutions which both satisfy inner
density condition. The following densities at the inner boundaries
were adopted: $\rho_{in}=10^{-12}{\rm \,g\,cm^{-3}}$ for $x_c=50$,
$\rho_{in}=10^{-14}{\rm \,g\,cm^{-3}}$ for $x_c=100$,
$\rho_{in}=10^{-16}{\rm \,g\,cm^{-3}}$ for $x_c=500$,
$\rho_{in}=10^{-21}{\rm \,g\,cm^{-3}}$ for $x_c=8000$.
The comparative results of the numerical integration of ~(\ref{dmless1})
for $s=1$ (GEF solution) and for $s=0$ (CAK wind theory) are shown
on Fig.1. These solutions were obtained for the following set of
parameters:

\begin{tabular}{|l|l|l|l|l|l|}
\multicolumn{4}{c}{ }\\
\hline $x_c^{\rm GEF}$ & $x_c^{\rm CAK}$ & $v^{\infty}_{\rm GEF}\,
\frac{\rm cm}{\rm s}$ &$v^{\infty}_{\rm CAK}\,\frac{\rm cm}{\rm s}$ & ${\dot
M}_{\rm GEF} \,\frac{\rm g}{\rm s}$ & ${\dot
M}_{\rm CAK} \,\frac{\rm g}{\rm s}$\\
\hline
$50$ & $75$ & $5\cdot 10^9$ & $3.67\cdot 10^9$  & $2.56597\cdot 10^{26}$ &
$2.55692\cdot 10^{26}$\\
\hline
$100$ & $155.2$ & $3.3\cdot 10^9$ & $2.6\cdot 10^9$ & $2.56597\cdot 10^{26}$ &
$2.55693\cdot 10^{26}$\\
\hline
$500$ & $751.7$ & $1.37\cdot 10^9$ & $1.15\cdot 10^9$ & $2.56599\cdot 10^{26}$ &
$2.55697\cdot 10^{26}$\\
\hline
\end{tabular}

\bigskip
\noindent
The difference between GEF and CAK wind is more pronounced when the
considerable portion of the wind is accelerated at a distance less than $100\,r_g$
from BH.
Note that the terminal velocity changes form $\Delta
v^{\infty}/v^{\infty}=0.36$ for $x_c=50$ to $\Delta
v^{\infty}/v^{\infty}=0.19$ for $x_c=500$.
The obtained results show that the GEF flow can
be sufficiently more fast than the flow which is described by CAK
theory.

\noindent
It is illustrative to demonstrate that CAK solution may
be obtained from ~(\ref{eqn1}) by "switching off" smoothly the
gravitational redshifting.
To obtain the continuous transition from the GEF solution to the CAK solution
of equation ~(\ref{eqn1}) the following procedure was adopted.
We assumed that the radiation force in ~(\ref{eqn1}) is
${\displaystyle f_L\sim dv/dr+\frac{1}{c}\,s\,GM/r^2}$
The introduced parameter $s$ continuously changes from
1 (GEF case) to 0 (CAK wind). Numerically calculating solutions of
~(\ref{eqn1}) for different values of $s$, applying the inner
density boundary conditions it is possible to demonstrate the
continuous transition of these solutions from the limiting cases of CAK
and GEF solutions. The results are shown on Figure 2.

The introduction of the Paczynski - Wiita ( PW ) ~(\ref{poten1}) potential, allows
to simulate the effects of general relativity. Equations analogous
to ~(\ref{dmless1}), ~(\ref{meq1}), ~(\ref{meq2}) are derived in Appendix. The
results of the numerical integration are shown on Figure 4.
Introduction of the modified potential can give a gain in $v^\infty$ that
varies from $\Delta v^\infty/v^\infty\sim 0.1$ for $x_c=30$ to
$\Delta v^\infty/v^\infty\sim 0.03$ for $x_c=100$.

\paragraph{Approximate relations for $a_1$, $p_c$.}
Introducing the following nondimmentional combination
${\displaystyle \epsilon=\frac{a}{c}}$,
equation ~(\ref{meq1}) will read:

\begin{equation}\label{meq11}
  4a_1 x_c^2 \epsilon p_c^2+(a_1^2-1)p_c-4a_1\epsilon=0\mbox{.}
\end{equation}
To obtain an approximate solution of ~(\ref{meq11}) we take into account that
$\epsilon<<1$. Equation ~(\ref{meq11}) has two roots:

\begin{equation}\label{rootp1}
(p_c)^{\pm}\simeq\frac{1}{8a_1\epsilon x_c^2}\left[1-a_1^2 \pm (1-a_1^2)\left(1
+\frac{32a_1^2 x_c^2\epsilon^2}{(a_1^2-1)^2}\right) \right]\mbox{.}
\end{equation}
From ~(\ref{rootp1}) we obtain

\begin{equation}\label{root_pm}
p_{c,1}\simeq-\frac{4a_1\epsilon}{1-a_1^2}\mbox{,}\qquad
p_{c,2}\simeq\frac{1-a_1^2}{4a_1 x_c^2\epsilon}\mbox{.}
\end{equation}

We are looking for the solution
with $p_c>0$ and with $a_1=a_1/v_{cr}<1$, thus the second root of ~(\ref{root_pm})
should be considered:

\begin{equation}\label{rootp11}
p\simeq\frac{1}{4a_1 x_c^2\epsilon}\mbox{,}
\end{equation}
where it was supposed that $a_1<<1$. The accuracy of
~(\ref{rootp11}) is $\Delta p_c/p_c \sim 10^{-3}$ (for $x_c=1$,
$p_c=12.0027$, $p^{appox}_1=12.0169$).

To obtain an approximate relation for $a_1$ we substitute $p_{c,2}$ from
 ~(\ref{root_pm}) to  ~(\ref{meq2}). Taking into account that $\beta=a_1/\epsilon$,
 the resulting equation reads:

\begin{equation}\label{roota1}
-2a_1^2 x_c \epsilon^2+\frac{\alpha-1}{\alpha}\frac{(1-a_1^2)^2}{4a_1}\epsilon+
\frac{a_1^2(1-\Gamma)}{2}-\frac{\epsilon a_1}{2\alpha}(1-a_1^2)=0\mbox{.}
\end{equation}
Neglecting in ~(\ref{roota1}) terms which contain $a_1^4$ and $a_1^3\epsilon^2$ will
result in the following equation for determining $a_1$:

\begin{equation}\label{roota11}
2a_1^3(1-\Gamma)-2a_1^2 \epsilon+\frac{\alpha-1}{\alpha}\epsilon=0\mbox{.}
\end{equation}
The second term in ~(\ref{roota11}) is small compared with other two. It allows to
obtain the following simple relation for $a_1$:

\begin{equation}\label{roota10}
a_1\simeq\left(\frac{\epsilon}{2(1-\Gamma)}\frac{1-\alpha}{\alpha}\right)^{1/3}\mbox{.}
\end{equation}
Calculating numerically $a_1$, $p_c$, from
~(\ref{meq1}),~(\ref{meq2}) we have found that the value of $a_1$ depends rather
weakly on $x_c$: $x_c=50$, $a_1=0.034347$, $p_c=48.0109$;
$x_c=500$, $a_1=0.034348$, $p_c=0.480112$; $x_c$=5000, $a_1=0.034369$, $p_c=0.004806$;
(obtained for $\alpha=0.6$, $\Gamma=0.5$).
Relation ~(\ref{roota10}) has a relatively good accuracy: for
$\epsilon=6\cdot 10^{-5}$ (\ref{roota10}) gives $a^{appr}_1\simeq
0.03430$ with accuracy
$\Delta a_1/a_1=10^{-3}$.
Velocity of the wind at the critical point is approximately $30\,
a_s$ ($v_{c}=a_1^{-1} a_s$).

Above the critical point the
influence of the $d\phi/dr$ (GEF) on
the additional acceleration
of the radiationally driven wind  is small compared to the effect due to $dv/dr$.

\section{Discussion}
The most successive model describing winds from hot stars
presented so far is that of Castor (1970),
Castor,Abbott \& Klein (1975). Geometry and ionization balance
differ O-star wind from AGN outflow. Line driven wind from O-type
star is spherically-symmetric. Outflows in AGN are
assumed to be originated from the luminous accretion discs and
approximately axially-symmetric. Powerful X-ray and UV
radiation from the disc pose a problem of overionization of the
outflowing plasma. In Sobolev approximation,
which is in the background of the CAK theory,
the radially streaming radiation flux is
absorbed in a line transition in a wind with the gradually
increasing velocity. A photon, emitted by a disc
will be red-shifted due to Doppler effect.
The resultant Sobolev optical depth $\tau_l\sim(dv/dr)^{-1}$.
As it was shown by CAK, the radiation force
that is due to an ensemble of optically thin and optically thick
lines is proportional to
$(dv/dr)^\alpha$.
The more $dv/dr$ the more effectively a line is shifted to the
powerful continuum.
In a number of studies, CAK theory
was applied to AGN in order to explain fast (up to $\sim 0.1 c$)
outflows Arav \& Li (1994), Arav, Li \& Begelman (1994),
Murray at al. (1995),
Proga et al. (1998), Proga et al. (2000).
The CAK theory was enhanced by adopting axial symmetry.
Ionization balance was studied in details simultaneously with
2D hydrodymanical calculations.

In this paper we developed a theory of winds that takes into
account effects of the strong gravitational fields. We point out
that if a wind is accelerated near super-massive BH a
gravitational change of photon's frequency must be taken into account. In a
strong gravitational field a photon, emitted by a disc, will be
red-shifted due to both the Doppler effect $\sim v/c$ and
gravitational redshifting $\Delta\nu/\nu=\Delta\phi/c^2$. We argue that
taking into account gravitational redshifting can substantially
change the wind dynamics and structure.
Although it should be mentioned that the developed theory (in the adopted toy model) cannot be
directly applied to explain outflows from close to BH.
An exact self-consistent solution of this problem is possible only in
general relativity. We avoid this
sophisticated task by considering all equations in flat space and time.

Basing on considerations
of Sobolev we conclude that the greater $d\phi/dr$ the more
effectively a line is shifted to the extent where the radiation
flux is unattenuated. In such a
case gravitational field "exposes" a wind to the unattenuated
continuum. Note that this effect is independent of the
wind dynamics, it works also in a medium with $v=0$.  We
determine a wind accelerated in this regime: "Gravitationally Exposed
Flow" (GEF).

\noindent The main goal of this paper has been to
compare a solution for GEF with that obtained from a standard line
driven wind theory ( CAK ). To solve this problem a very simple
input physics was assumed: spherical symmetry, constant
temperature  and no ionization balance. We found that in such a
case the radiation force
$f_l \sim(dv/dr+\frac{1}{c}d\phi/dr)^{-1}$.
Numerical analysis shows that the introduced effect changes drastically a
slope of a solution curve at the bottom of the wind. It is clearly
a result of the fact that the role of the gravitational field is
important when the field is high an velocity gradient is low.
As soon as $dv/dr$ becomes
sufficient, Sobolev effect efficiently stimulates the acceleration
of the wind.
Obtained equation of motion is nonlinear with respect to $dv/dr$
but this nonlinearity is different form that of CAK.
Critical point of the equation of motion is neither a sonic point
nor a CAK critical point. The position of the critical point $r_{c}$ is
the only free parameter of the problem. Adjusting $r_{c}$ in
order to satisfy density boundary conditions we compared solution
for GEF with that of CAK which was obtained for the same physical and
boundary conditions.
Numerically solving equations describing
spherical symmetric, stationary outflowing wind we found that
gravitational redshifting can make
acceleration up to $35\%$ (for $r_{cr}=50\,r_g$) more efficient.

In order to take approximately into account effects of general
relativity, we made use of modified potential (Paczynski-Wiita potential).
Numerical analysis demonstrated that if the critical point is located as far as
$30\,r_g\div 100\,r_g$ the difference between PW case and Newtonian case is $\Delta
v^{\infty}/v^{\infty}=0.1\div 0.03$

\bigskip
\noindent
{\it\bf Acknowledgements }
Author is deeply grateful to G.S. Bisnovatyi-Kogan for continuous
and stimulating attention to this work.

\section{Conclusions}
\begin{enumerate}
\item
A theory of winds accelerated by the radiation
pressure in lines with account of gravitational redshifting of
photons is developed. A system of equations describing stationary,
spherically-symmetric, isothermal flow is derived.

\item
A solution of these equations is obtained numerically for two cases:
a standard line-driven wind (CAK theory), Gravitationally Exposed Flow (GEF)-
a wind that is accelerated by the radiation pressure in lines
if to take
into account gravitational redshifting. It is shown that an increase of up to
$35\%$ in $v^\infty$ can be obtained.

\item
To take approximately into account effects of general relativity,
Paczynski - Wiita potential is adopted. A wind solution is
calculated for this type of potential and comparative analysis with
Newtonian case is presented.

\item
The developed theory can be used to explain fast outflows
from AGN.
\end{enumerate}

\begin{description}
\item
Abbott D.
1980, {\it ApJ}, 242, 1183

\item
Arav N., Korista, K.T., Martijn de Kool
2002, {\it ApJ}, 566, 699

\item
Arav N., Li Z.Y.
1994, {\it ApJ}, 427, 700

\item
Arav N., Li Z.Y., Begelman M.C.
1994, {\it ApJ}, 432, 62

\item
Artemova I.V., Bjornsson G., Novikov I.D. 1996,
{\it ApJ}, 456, 119a

\item
Bisnovatyi-Kogan G.S., Dorodnitsyn A.V. 1999
{\it A\&A}, 344, 647

\item
Bisnovatyi-Kogan G.S. 2001,
Stellar Physics, Vol.2, Springer, 2001

\item
Castor J.I.
1970, {\it MNRAS}, 149, 111

\item
Castor J.I., Abbott D.C, Klein R.
1975, {\it ApJ}, 195, 157

\item
Landau L.D., Lifshitz E.M. 1960,
The Classical Theory of Fields,
New York: Pergamon

\item
Lucy L.B., Solomon P.
1970, {\it ApJ}, 159, 879

\item
Murray N., Chiang J., Grossman S.A., Voit G.M.
1995, {\it ApJ}, 451, 498

\item
Proga D., Stone J.M., Drew J.E.
1998, {\it MNRAS}, 295, 595

\item
Proga D., Stone J.M., Kallman T.R.
2000, {\it ApJ}, 543, 686

\item
Stevens I.R., Kallman T.R.
1990, {\it ApJ}, 365, 321

\item
Sobolev V.V. 1960,
Moving envelopes of stars,
Cambridge: Harvard University Press

\item
Tennant, A.F., Mushotzky, R.F., Boldt, E.A. and Swank, J.H.
1981, {\it ApJ}, 251, 15

\end{description}

\newpage
\section{Appendix}
Substituting Paczynski - Wiita potential into equation
~(\ref{momentum1}) will obtain equation of motion:

\begin{eqnarray}
F&=&\left(1-\frac{a}{v^2}\right)v\frac{dv}{dr}-\frac{2a^2}{r}+
\frac{GM}{(r-r_g)^2}-\frac{L\sigma}{4\pi r^2 c}\nonumber\\
&-&\frac{L\sigma k}{4\pi r^2 c}
\left(\frac{4\pi}{|{\dot M}|\sigma
v_{th}}\right)^\alpha\left[vr^2\left(\frac{dv}{dr}+\frac{1}{c}
\frac{GM}{(r-r_g)^2}\right)\right]^\alpha=0\mbox{,}\label{eqn1PW}
\end{eqnarray}
Scaling all variables according to ~(\ref{nondimvar}), omiting
tilda, equation ~(\ref{eqn1PW}) reads:

\begin{eqnarray}
&&F(p,v,x)=\left(1-\frac{a^2_1}{v^2}\right)v p x^2 -2a_1^2 x+
\frac{\beta^2}{2}\left(\frac{x^2}{(x-1)^2}-\Gamma\right)\nonumber\\
&&-\mu\beta^2\left[\frac{v}{\beta^2}\left(p+
\frac{\beta}{2(x-1)^2}\right)\right]^\alpha=0\mbox{,}\label{PW1}
\end{eqnarray}
where $p\equiv dv/dx$, $\mu$ is determined by ~(\ref{mu1}) and $a_1$, $\beta$ by
~(\ref{param}).
Making use of ~(\ref{singcond}) and ~(\ref{regcond}) will result
in the following relations:

\begin{equation}\label{meq1PW}
\frac{\alpha-1}{\alpha} (1-a_1^2)x_c^2 p_c-2a_1^2 x_c
+\frac{\beta^2}{2}(\frac{x_c^2}{(x_c-1)^2}-\Gamma)-
(1-a_1^2)\frac{\beta}{2\alpha}\frac{x_c^2}{(x_c-1)^2}=0\mbox{,}
\end{equation}

\begin{equation}\label{meq2PW}
c_2 p_c^2+c_1 p_c+c_0=0\mbox{.}
\end{equation}
Coefficients $c_i$ are determined from the following relations:

\begin{eqnarray}\label{ci}
c_2&=&2a_1^2 x_c^2\mbox{,}\\
c_1&=&\frac{(a_1^2-1)x_c^2 \beta}{2(x_c-1)^2}\mbox{,}\\
c_0&=&\frac{-x_c(\beta-1)\beta + a_1^2(-2(x_c-1)^3-\beta x_c)}
{(x_c-1)^3}\mbox{.}
\end{eqnarray}

Equation for $\mu$ reads:
\begin{equation}\label{mu2}
\mu=\alpha^{-1} (1-a_1^2)\left[\frac{x_c^2}{\beta^2}\left(p_c
+\frac{\beta}{2(x_c-1)^2}\right)\right]^{1-\alpha}\mbox{.}
\end{equation}
Equation ~(\ref{mu2}) is analoguis to ~(\ref{mu}) except the last
term in brackets. For a given $x_c$, equations ~(\ref{meq1PW}),~(\ref{meq2PW})
are solved numerically to determine $p_c$ and $a_1$. Then
~(\ref{PW1}) is integrated numerically as described in the text.
Comparative results of the numerical solution of ~(\ref{PW1}) are
depicted of Figure 4.

\begin{figure}
\centerline{\psfig{figure=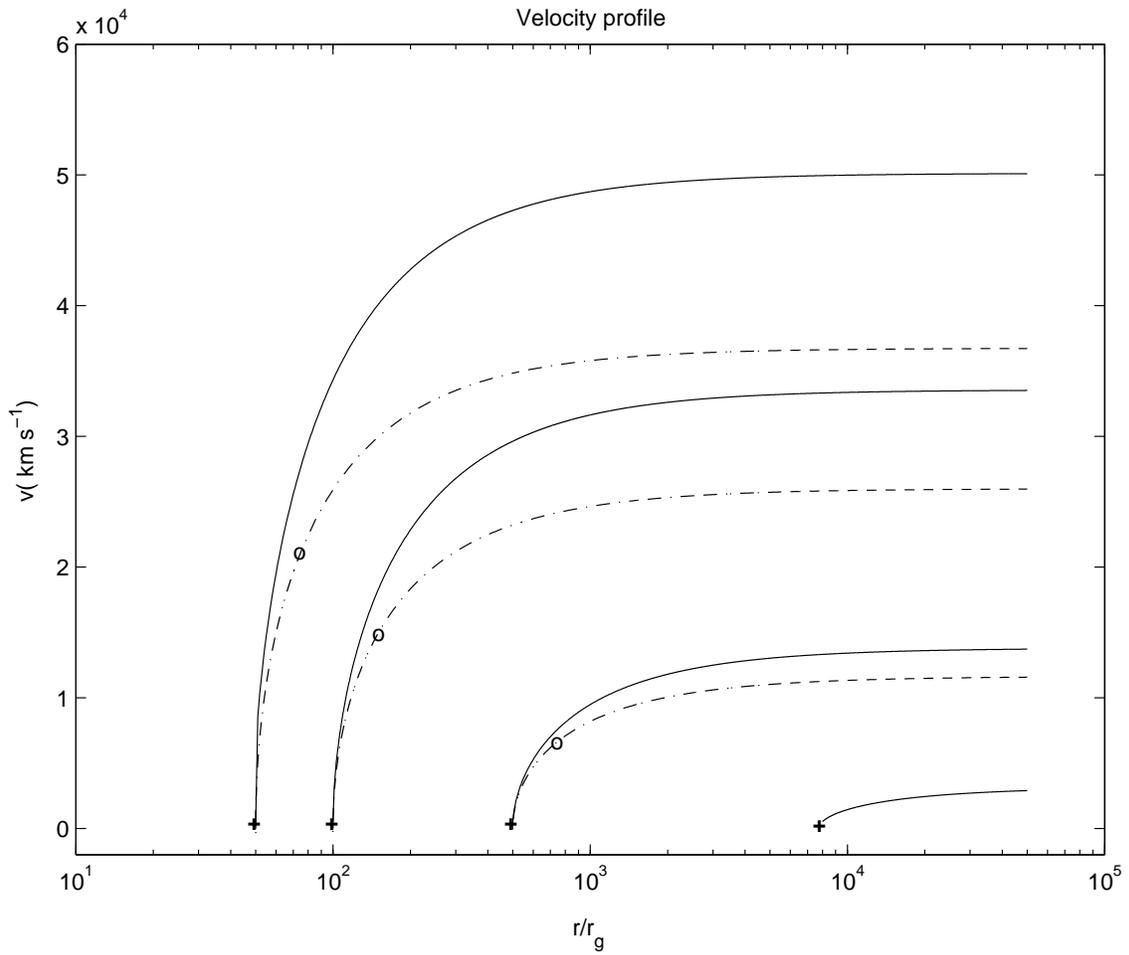,height=5in}}
\caption{Solution of the wind equation ~(\ref{dmless1}).
Solid line - GEF solution, dashed line - CAK solution. Crosses
indicate GEF critical points, circles - CAK critical points.
Curves for $x=8000$ are graphically indistinguishable.}
\end{figure}

\begin{figure}
\centerline{\psfig{figure=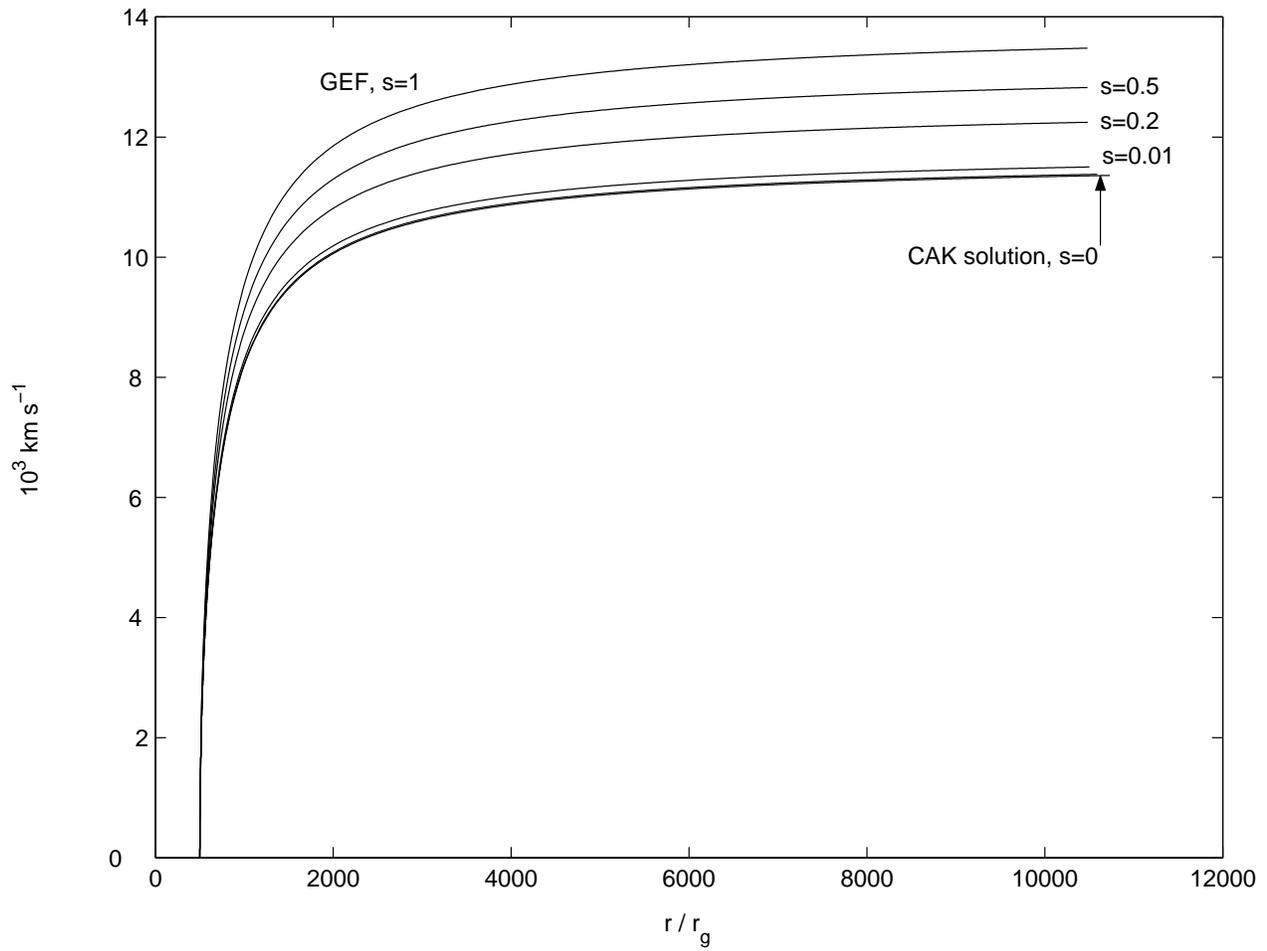,height=5in}}
\caption{The continuous transition of the solution of ~(\ref{dmless1}) from
GEF (s=1) to CAK regime (s=0), see explanations in the text.
}
\end{figure}

\begin{figure}
\centerline{\psfig{figure=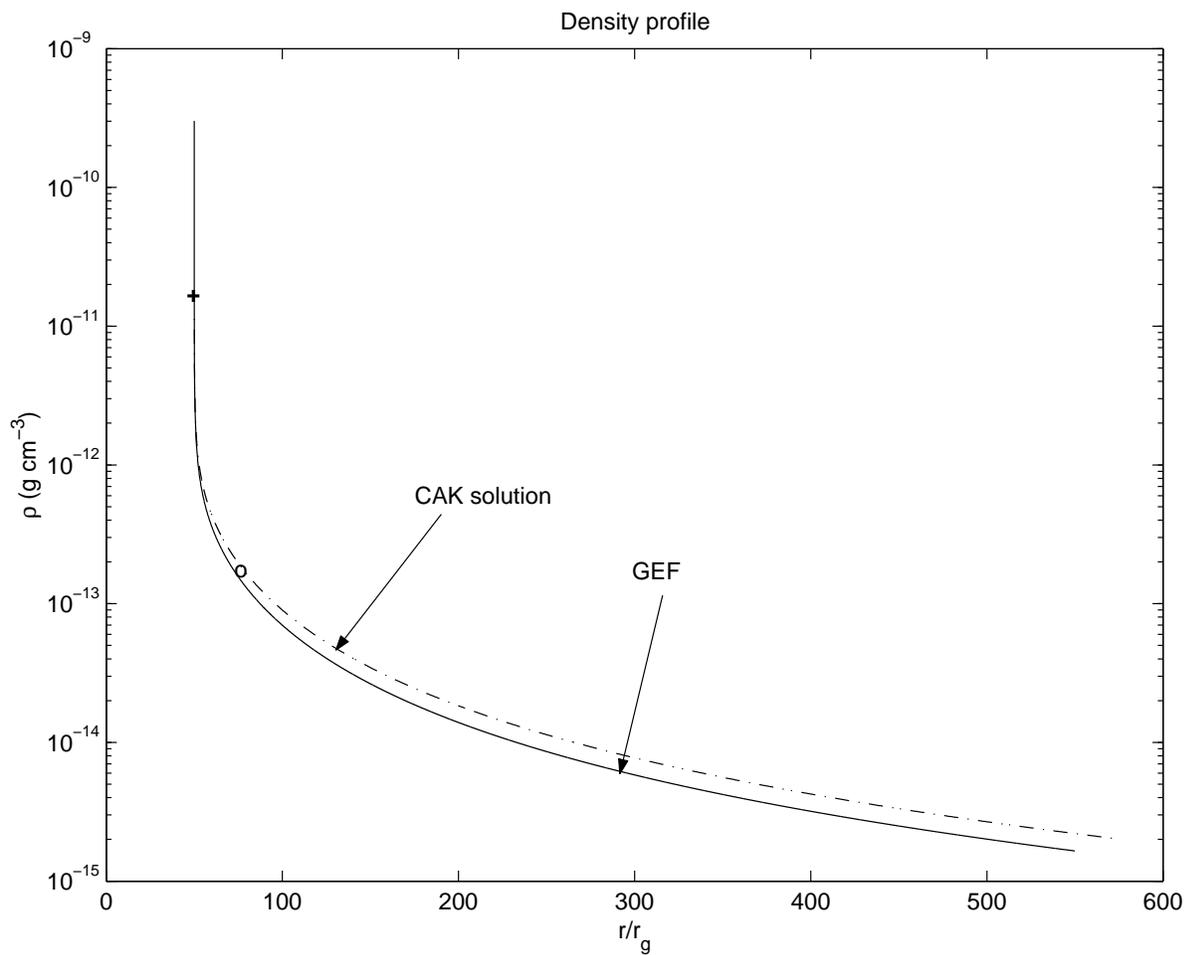,height=5in}}
\caption{Density distribution in the wind. The solution curves
calculated for $x^{GEF}_c=50$, $x^{GEF}_c=75$.}
\end{figure}

\begin{figure}
\centerline{\psfig{figure=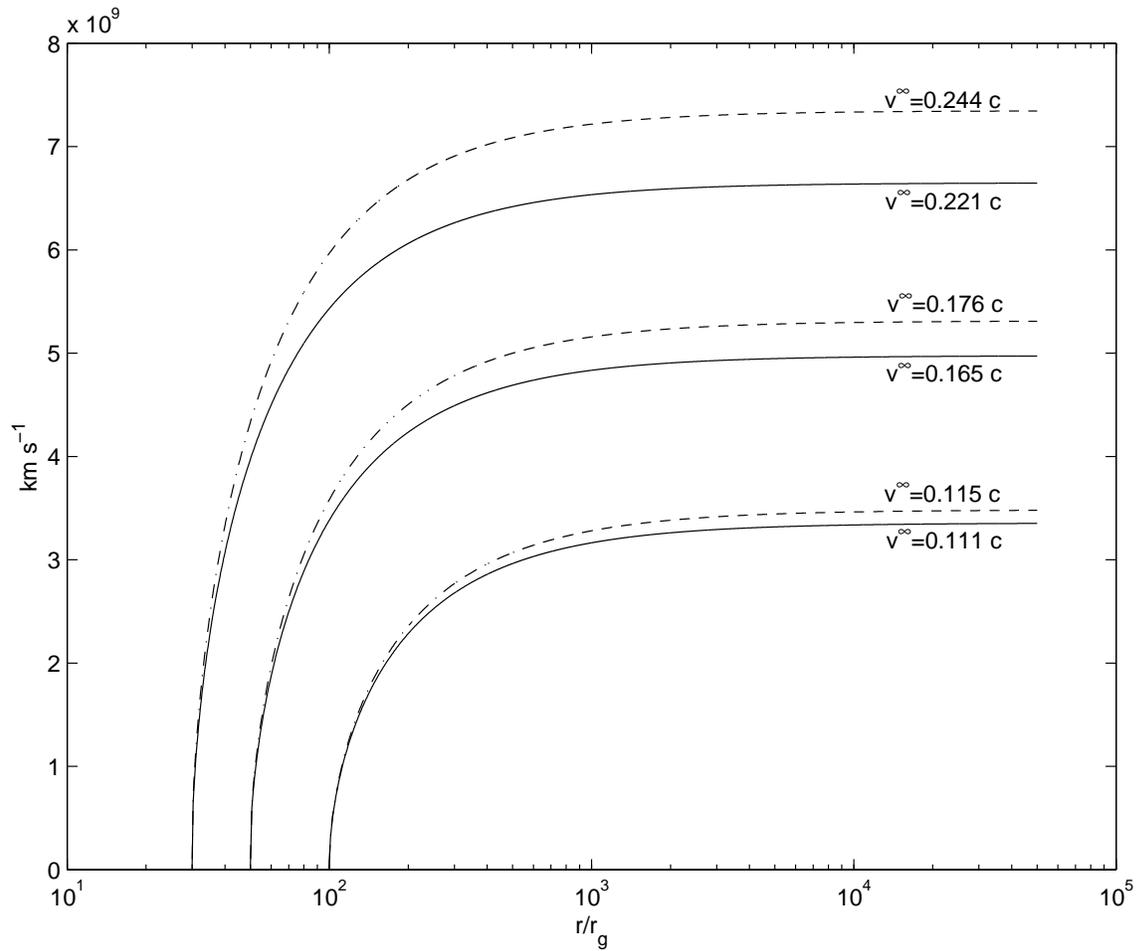,height=5in}}
\caption{
Solid line indicates GEF solution for the Newtonian potential,
dashed line GEF solution for Paczynski - Wiita potential.
}
\end{figure}

\end{document}